\documentclass[12pt]{article}
\usepackage{epsfig}

\usepackage{amssymb}

\def\be{\begin{equation}}
\def\ee{\end{equation}}
\def\ba{\begin{array}{c}}
\def\ea{\end{array}}

\def\ben{$$}
\def\een{$$}

\def\bea{\begin{eqnarray}}
\def\eea{\end{eqnarray}}

\def\beax{\begin{eqnarray*}}
\def\eeax{\end{eqnarray*}}

\begin{document}


\vspace{.35cm}

 \begin{center}{\Large \bf

The large$-g$ observability of the low-lying energies  in the
strongly singular potentials $V(x)=x^2+g^2/x^6$  after their ${\cal
PT}-$symmetric regularization

  }\end{center}

\vspace{10mm}

 \begin{center}

 {\bf Miloslav Znojil}

 \vspace{3mm}
Nuclear Physics Institute ASCR, 250 68 \v{R}e\v{z}, Czech Republic

{e-mail: znojil@ujf.cas.cz}


%
\vspace{3mm}

%

%
%
%
%
%


\vspace{13mm}

\end{center}

\vspace{5mm}


\subsection*{Abstract}

The elementary  quadratic plus inverse sextic interaction $V(x)=x^2+
g^2/x^6$ containing a  strongly singular repulsive core in the
origin is made regular by a complex shift of coordinate $x = s-{\rm
i}\varepsilon$. The shift $\varepsilon>0$ is fixed while the value
of $s$ is kept real and potentially observable, $s \in
(-\infty,\infty)$. The low-lying energies of bound states are found
in closed form for the large couplings $g \gg 1$. Within the
asymptotically vanishing ${\cal O}(g^{-1/4})$ error bars these
energies are real so that the time-evolution of the system may be
expected unitary in an {\em ad hoc} physical Hilbert space.

 \noindent

\subsection*{Keywords}

quantum evolution; triple-Hilbert-space picture; strongly singular
forces; regularization by complexification; strong-coupling
dynamical regime; unitarity;

\newpage

\section{Introduction \label{intr} }

Although the differences between the classical and quantum laws of
evolution are deep, the search for their practical applications
often proceeds in unexpectedly close parallels. For illustration one
may recall the existence of certain parallelism between the
principles and concepts of classical and quantum computing. One of
the important ingredients in these developments may be seen in the
recent thorough changes in the perception of quantum systems and of
their theoretical description. In place of the traditional idea of
preparation of a quantum wave function $\psi$ at time $t_0=0$ and of
its exposition to measurement at time $t_1>0$, several innovations
of the paradigm have emerged. In our present paper, specific
attention will be paid to the underlying problem of the prediction
of the results of the {\em unitary} evolution, based on the more or
less routine solution of Schr\"{o}dinger equation
 \be
 {\rm i}\partial_t\,\psi = H\,\psi\,
 \ee
in which the Hamiltonian $H$ is NOT assumed self-adjoint, $H \neq
H^\dagger$.

Review papers \cite{Geyer,Carl,ali,SIGMA} may be recalled for an
exhaustive introduction into the underlying version  of quantum
theory in which the wave function $\psi$ is perceived represented,
in parallel, in three alternative Hilbert spaces ${\cal
H}^{(F,S,T)}$ where just ${\cal H}^{(S)}$ and ${\cal H}^{(T)}$ are
assumed unitarily equivalent. The respective superscripts may be
read as abbreviating ``first'', ``'second'' and ``third'' Hilbert
space. The ``first'' space is most important as being, in the
textbooks, ``favored'' as the ``friendliest'' one.

Whenever one restricts attention to the most common one-dimensional
kinetic plus local potential Hamiltonians $H=-d^2/ds^2 +V(s)$, there
rarely emerge any doubts about the choice of the first space in the
form of the Hilbert space of the square-integrable  functions of
$s$, ${\cal H}^{(F)} \ \equiv \ L^2(\mathbb{R})$. In this context,
Bender with Boettcher \cite{BB} were probably the first quantum
theoreticians who sufficiently explicitly emphasized that such a
choice of the representation space  may be wrong and unphysical. In
other words, the above reading of the superscript may happen to
change into ``false'' \cite{SIGMA}.

Extremely persuasively, Bender and Boettcher illustrated the
validity of such a claim by the demonstration of the reality of the
spectrum of bound states (and, hence, of the possible unitarity of
the quantum evolution after an {\em ad hoc} amendment ${\cal
H}^{(F)}\to {\cal H}^{(S)}$ of the physical Hilbert space of states)
even when generated by the following, {\em manifestly non-Hermitian}
confining potential
 \ben
 V^{(BB)}(x)= ({\rm i}x)^{\delta}\,x^{2}\,,\ \ \ \
 \delta \geq 0\,.
 \een
We feel inspired by such a claim. We intend to extend the scope of
the corresponding mathematics as summarized, e.g., in review paper
\cite{ali} to the domain of applications in which the non-Hermitian
confining potentials are allowed strongly singular.

For the sake of definiteness we shall pay our attention to the
potentials exhibiting not only the left-right symmetry of the
problem in the complex plane (also known as ${\cal PT}-$symmetry
\cite{Carl}) but also the following strong form of the
central-repulsion property,
 \be
 V(x)= g^2\,x^{-2-\gamma}+ {\rm subdominant \ terms} \,,\ \ \ \ \
 \gamma \geq 0\,,\ \ \ \ \
 |x| \ll 1\,.
 \label{crp}
 \ee
Our decision and project of study were motivated by the challenging
available results as obtained, say, in refs.~\cite{BG,regul} for the
$\gamma=0$ weaker-singularity cases. One should add that at
$\gamma=0$ the regularization proved comparatively easy as long as
it could rely upon the centrifugal-force nature of the  $\gamma=0$
repulsion. At the same time, one may expect that at $\gamma>0$ and
$g^2>0$ the regularization of the singular spike may still be
achieved by the same (in fact, by the Buslaev's and Grecchi's
\cite{BG}) complex shift of the coordinate $x = s-{\rm
i}\varepsilon$ in which the real shift parameter $\varepsilon>0$ is
a constant while the value of $s$ is kept variable, $s \in
(-\infty,\infty)$.

The text of our paper will be separated in section \ref{dva} (in
which a specific model is chosen for our study), section \ref{tri}
(in which we evaluate the low lying spectrum via a suitable
perturbation technique) and discussions (section \ref{ctyri}).

\section{The model\label{dva}}

The studies of the strongly spiked repulsive interactions
(\ref{crp}) usually find motivation in perturbation theory
\cite{Harrell} and in computational physics \cite{numerical} as well
as in field theory \cite{Lukyanov} and in descriptive and
phenomenological contexts \cite{Klauder}. In our present paper
devoted to the study of exactly solvable extremes we shall pick up
one of the most elementary interactions
 \be
  V(x)= x^2 + \frac{g^2}{x^6}\,
  \label{pot}
  \ee
where $\gamma=4$ and where the real coupling $g^2$ will be
considered very large, $g \gg 1$. In addition, the related ordinary
differential Schr\"{o}dinger equation
 \be
 \left [-\frac{d^2}{dx^2}+V(x)
 \right ]\,\psi_n(x)=E_n\,\psi_n(x)
 \,
 \label{SEst}
 \ee
will be defined along the straight complex line of $x=x(s) \in
\mathbb{C}$ where the parameter $s$ runs over the whole real domain,
$ s \in \mathbb{R}$. The distance $\varepsilon > 0$ of this line
from the real axis of $x$ will be kept fixed,
 \be
  x = x(s)= s - {\rm i}\varepsilon\,.
  \label{BGpath}
 \ee
The most common square-integrable bound-state solutions of our
Schr\"{o}dinger Eq.~(\ref{SEst}) in ${\cal H}^{(F)} \ \equiv \
L^2(\mathbb{R})$ will be required to satisfy the usual Dirichlet
asymptotic boundary conditions,
 \be
 \psi_n(x)=\psi_n[x(s)]=\phi_n(s) \to 0 \ \ \ {\rm for} \ \ \ s \to
 \pm
 \infty\,.
 \label{bc}
 \ee
Finally, the standard probabilistic interpretation of these states
will be assumed achieved, in principle at least, via a change of the
inner product in ${\cal H}^{(F)}$ (yielding a ``standard'' physical
Hilbert space ${\cal H}^{(S)}$) or, if asked for, via a subsequent
replacement of the second space ${\cal H}^{(S)}$ by its unitarily
equivalent alternative ${\cal H}^{(T)}$. For more details the
readers may check, e.g., ref.~\cite{SIGMA} in which we worked with
superscript $^{(P)}$ (marking, in the inspiring context of
ref.~\cite{Geyer}, the ``primary'' physical space) in place of the
present, less sophisticated superscript $^{(T)}$ meaning just the
``third'' space.

The analyticity of our potential $V(x)$ in the complex plane of $x$
(with the exemption of the origin and infinity) implies immediately
\cite{Sibuya} that the wave-function solutions $\psi_n(x)$ of the
ordinary differential Eq.~(\ref{SEst}) may be considered analytic in
the complex plane of $x$, endowed with a properly chosen cut which
would connect the origin with infinity. Naturally, for $\varepsilon
> 0$ we shall choose the cut starting at the origin and oriented
upwards.
%
%
%
%

From the purely phenomenological point of view one of the key merits
of our model (\ref{SEst}) + (\ref{BGpath}) may be seen in our
freedom of choosing {\em any}  parameter $\varepsilon > 0$. In the
spirit of the standard oscillation theorems which hold for the
complex linear differential equations of the second order
\cite{Ende} we know that due to the analyticity of the wave
functions below the real line of $x$ the variations of the value of
$\varepsilon>0$ will leave the spectrum unchanged. Thus, in
particular, the choice of a very small $\varepsilon$ will enable us
to approximate some of the ``standard textbook'' $\varepsilon=0$
bound-state wave functions by the not too different related
solutions $\psi_n(x)$ of our present ``regularized'',
$\varepsilon\approx 0$ eigenvalue problem.

In contrast,  the choice of a very large $\varepsilon$ will simplify
the mathematics. We shall show below that the independence of the
spectrum of the value of the shift parameter $\varepsilon$ may
enable one to combine a useful predictive power of the model with
certain friendly and constructive mathematical features.

%
%
%
%
%
%

\section{Strong-coupling perturbation expansions\label{tri}}


In our Schr\"{o}dinger Eq.~(\ref{SEst}) the Hamiltonian is evidently
non-Hermitian in $L^2(\mathbb{R})$ where, via Eq.~(\ref{BGpath}),
the symbol $\mathbb{R}$ represents the real line of variable $s$.
This Hilbert space must be declared unphysical and ``false'',
therefore \cite{SIGMA}. A deeper explanation of such an apparent
paradox may be found in Refs.~\cite{Geyer,Carl,Carlb}. Its
resolution is easy, in principle at least. An {\em ad hoc}
redefinition of the inner products in $L^2(\mathbb{R})$ does the
job. It makes the Hamiltonian, by construction, self-adjoint in the
new, Hamiltonian-adapted Hilbert space ${\cal H}^{(S)}$ where the
superscript $^{(S)}$ abbreviates the word ``standard''~\cite{SIGMA}.

Let us now  return to the approximate solution of our differential
Schr\"{o}dinger eigenvalue problem (\ref{SEst}) + (\ref{BGpath}) +
(\ref{bc}), i.e., to the ordinary differential equation
 \be
 \left [-\frac{d^2}{ds^2}+W(s)
 \right ]\,\phi_n(s)=E_n\,\phi_n(s)
 \,,\ \ \ \ \ \
 W(s)\ \equiv \ V[x(s)]=
 (s - {\rm i}\varepsilon)^2 + \frac{g^2}{(s - {\rm i}\varepsilon)^6}\,
 \label{SEdum}
 \ee
for the bound-state wave functions re-written in the equivalent form
$\phi_n(s) \in L^2(\mathbb{R})$ and living on the real line of $s$.
We shall restrict our attention to the strong-coupling dynamical
regime where $g \gg 1$. We shall see that in such a dynamical regime
one is permitted to use certain less usual perturbation expansions
in a small parameter $1/g^{c}$ such that $0< c < 1$.


The detailed description of the method may be found in
Ref.~\cite{Omar} where the perturbation recipe has successfully been
tested and found to work for complex-valued potentials. Here, it is
only necessary to demonstrate that the presence of the repulsive
barrier will not obstruct the applicability of such a
perturbation-expansion technique.

\subsection{Approximations using special values of $\varepsilon$}

\begin{figure}[h]                     
\begin{center}                         
\epsfig{file=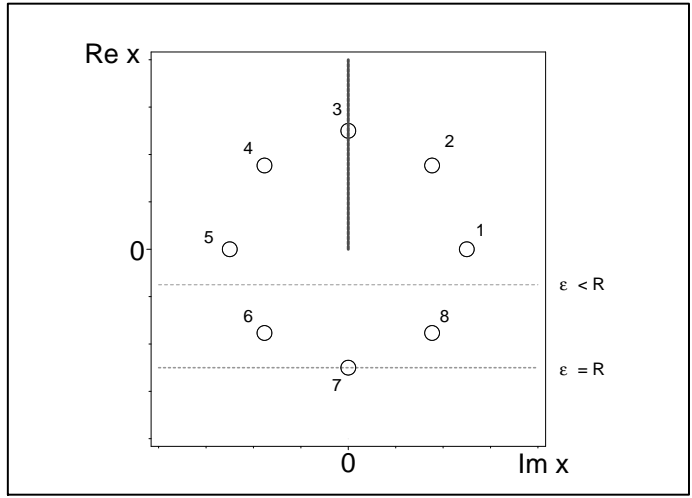,angle=270,width=0.6\textwidth}
\end{center}                         
\vspace{-2mm} \caption{Positions of the eight roots $x=R_m$ (small
circles) in complex plane of $x$ (with the thick-line cut oriented
upwards). Two horizontal lines sample the eligible integration paths
(\ref{BGpath}). The lower one which crosses the stationary point
$R_7$ is exceptional, yielding the feasible single-well
harmonic-oscillator approximation. \label{firmen}}
\end{figure}

Perturbation theory as explained, e.g., in Ref.~\cite{Omar} tells us
that its convergence to exact results cannot be guaranteed in
general. Thus, one only has to use the formalism as a source of
suitable asymptotic series and approximants.
%
%
In this sense, the whole perturbation recipe will satisfy our
present needs. Its essence can be summarized as based on several
assumptions. Firstly,
%
%
%
%
%
%
%
we shall require that the first derivative of our complex potential
function will vanish at a certain complex value of $x=R_m$. This is
our first stationarity requirement which reads $V'(R_m)=0$ and
yields an elementary algebraic equation for all of the eligible
complex points $R_m$,
 \ben
 2R_m = \frac{6g^2}{R^7_m}\,.
 \een
Such an algebraic equation possesses eight well-separated
closed-form complex roots at which our potential is locally
constant,
 \ben
 R_m=R\,e^{ {\rm i}\,\pi\,(m-1)/4}\,,\ \ \ \
 R=|3^{1/8}g^{1/4}|\gg 1\,,\ \ \ \
 m =
 1,2,\ldots,8\,.
 \een
In Fig.~\ref{firmen} the positions of the roots $x=R_m$ in complex
plane of $x$ are marked by the small numbered circles.

The well known second stationarity condition is the positivity of
the second derivative of our potential $V(x)$ at the eligible
stationary point $x=R_m$. Fortunately, irrespectively of the
subscript $m$ we obtain the same real and positive value of this
derivative,
 \ben
 V''(R_m)= 2+42\,\frac{g^2}{R^8_m}= 2+42\,\frac{g^2}{3g^2}=16\,.
 \een
Thus, all of the eight points of stationarity are equally friendly
from this point of view.

\begin{figure}[h]                     
\begin{center}                         
\epsfig{file=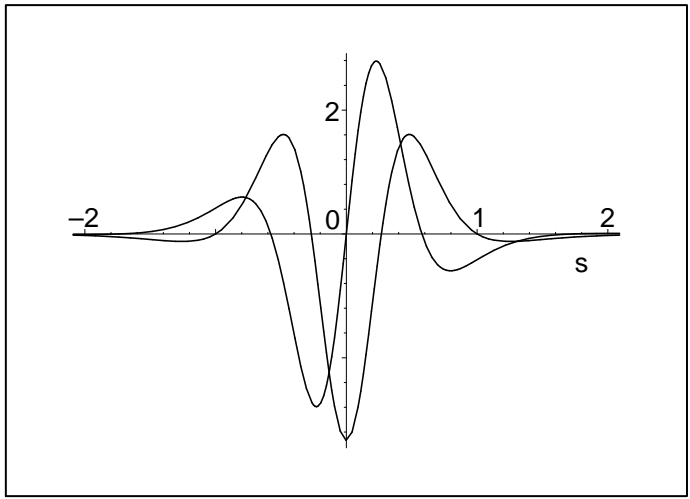,angle=270,width=0.6\textwidth}
\end{center}                         
\vspace{-2mm} \caption{The rescaled real part ($\varrho^{-1}
Re\{V[x(s)]\}$, symmetric curve) and rescaled imaginary part
($\varrho^{-1} Im\{V[x(s)]\}$, antisymmetric curve)  of potential of
Eq.~(\ref{potuj}) at $\varepsilon=R/100=1$ using the huge rescaling
factor $\varrho = 10^{15}$. \label{firm}}
\end{figure}

In the next step of our analysis we are free to select an optimal
value of the optional complex shift $\varepsilon>0$ in integration
path of Eq.~(\ref{BGpath}). In Fig.~\ref{firmen} the thinner, upper
horizontal line samples the generic case using a not too large shift
$\varepsilon<R$. Along this line the shape of our complex potential
function $V[x(s)]$ remains complicated and unfriendly to
approximations (see its typical sample as presented in
Fig.~\ref{firm}).

%

\begin{figure}[h]                     
\begin{center}                         
\epsfig{file=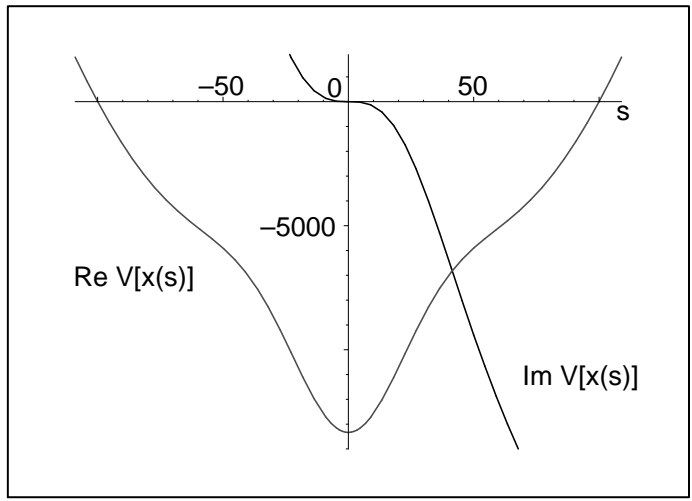,angle=270,width=0.6\textwidth}
\end{center}                         
\vspace{-2mm} \caption{The real (symmetric) and imaginary
(antisymmetric) parts of potential $V[x(s)]$ of Eq.~(\ref{potuj}) at
$\varepsilon=R=100$. \label{firmonej}}
\end{figure}

The thicker, lower horizontal line of Fig.~\ref{firmen} is chosen as
crossing the seventh stationary point. The related shape of our
complex potential function $V[x(s)]$ in sampled in
Fig.~\ref{firmonej}. We immediately see that after such a special
choice of $\varepsilon=R$, the shape of our potential $V[x(s)]$
resembles, very closely, the spatially symmetric harmonic-oscillator
well near its minimum, with a very small admixture of an asymmetric
imaginary component. Thus, an approximation of our interaction by a
confining effective harmonic-oscillator potential may be expected
very efficient at $\varepsilon=R$.

A detailed constructive proof of the latter expectation will be
delivered in the next subsection. Before moving to this proof, let
us briefly return to the other possibilities of the special choices
of $\varepsilon<R$. First of all, we may exclude the use of the
non-positive shifts $\varepsilon\leq 0$ as highly uncomfortable
because the corresponding integration path would have to cross the
cut or singularity.

In this sense the only remaining and apparently user-friendly
alternative would be the choice of the integration line
(\ref{BGpath}) with $\varepsilon=R/\sqrt{2}$. It would pass strictly
through the {\em pair} of the left and right stationary points $R_6
= x(-1/\sqrt{2})$ and $R_8 = x(+1/\sqrt{2})$, respectively.
Naturally, the resulting double-well problem (in which the two
far-away wells are equally deep) is enormously complicated. In any
case, it would not admit any easy perturbative treatment. Interested
readers may find a deeper related methodical analysis in
Ref.~\cite{Jakub}.


\subsection{Harmonic-oscillator approximation}

Once we choose the path of Eq.~(\ref{BGpath}) with $\varepsilon =R$
which crosses the only eligible candidate $R_7$ for a ``useful''
stationary point, we come to the conclusion that we satisfied all
requirements and that we may finally apply the recipe of
Ref.~\cite{Omar}. Let us now describe the results in full detail. In
the first step let us reparametrize our potential,
 \be
  V(x)= x^2 + \frac{R^8}{3x^6}\,
  \label{potuj}
  \ee
and note that at large real $R$ and at the small complex shifts
${\rm i}\varepsilon= s - x(s)$ the shape and $s-$dependence of
function (\ref{potuj}) is wild and dominated by its singular part
(cf. Fig.~\ref{firm}). At the large values of $\varepsilon \sim R$
the situation is different.  The value of the complex function
$V[x(s)]$ if dominated by its real part. Moreover, the latter real
function of $s$ has a deep minimum at $x=-{\rm i}R$ (cf.
Fig.~\ref{firmonej}). In such a case it makes sense to expand our
complex potential near the stationary point $R_7 \equiv -{\rm i}R$
in Taylor series. This yields the following complex power series in
real variable $s$,
 \ben
 V[x(s)]
 = -\frac{4}{3}{R}^{2}+8{s}^{2}-{\rm i}{\frac {56}{3R}}{s}^{3}-
 \frac{42}{R^2}{s}^{4}+{\rm i}{\frac
 {84}{{R}^{3}}}{s}^{5}+\frac{154}{{R}^{4}}{s}^{6}-
 \een
 \be
 -{\rm i}{ \frac
 {264}{{R}^{5}}}{s}^{7}-\frac{429}{{R}^{6}}{s}^{8}+{\rm i} \frac {
 2002}{3{R}^{7}}{s}^{9}+
 \frac{1001}{{R}^{8}}{s}^{10}-\ldots\,.
 \label{taylor}
 \ee
The radius of convergence of the series is  equal to $R$ and also
the typical numerical factors of suppression of the next-order term
remain large. Incidentally, the sequence of the polynomial
truncations of the Taylor series may be found equal to the sequence
of  popular power-law interactions exhibiting ${\cal PT}-$symmetry
\cite{Carl}. Thus, we may truncate this series an insert the
resulting polynomial interaction in our differential Schr\"{o}dinger
equation.

In the initial step of the construction we restrict our attention to
the first two terms of series Eq.~(\ref{taylor}) and arrive at the
exactly solvable model of the usual, real harmonic oscillator.
%
%
This enables us to identify the low-lying spectrum of bound states
of our model with the {\em real} energies given by the following
closed formula,
 \be
 E_n=-\frac{4{R}^{2}}{3}+(2n+1)\sqrt{8} + {\cal O}\left (
 \frac{1}{R}
 \right )\,,
 \ \ \ \ \ n = 0, 1, \ldots\,.
 \ee
Finally, it is quite routine to show that all of the higher-order
corrections to the potential in series Eq.~(\ref{taylor}) lead to
the asymptotically vanishing corrections to the energies. We may
summarize that within the limits of the first few orders of
perturbation theory our model remains solvable. It is also worth
noticing that the low-lying energy levels are perceivably negative
and approximately equidistant.

\section{Discussion\label{ctyri}}

Certainly, our present result is perturbative so that the problem of
the rigorous proof of the strict reality of the spectrum remains
open. At the same time, the spectrum of energies of our present
model {\em is} real within the precision offered by the perturbation
series. The validity of such an observation may be further supported
when we notice that the integrals representing the contribution of
the odd powers in series (\ref{taylor}) (i.e., of the purely
imaginary corrections threatening to introduce also an imaginary
component into the spectrum) vanish identically.

Naturally, whenever we are interested in the exact, non-perturbative
energies our present arguments  remain inconclusive (cf. also a few
general relevant remarks on this topic in \cite{Alvarez}). Our
observation that the imaginary odd powers of $s$ in (\ref{taylor})
cannot contribute to the Rayleigh-Schr\"{o}dinger series only
supports the reality of the spectrum in the limit $R \to \infty$.

Let us add that along the complex line of $x(s) = s-{\rm i}R$ also
the analytic wave functions $\psi(x)$ of our model will coincide,
with very reasonable precision, with the well known
Hermite-polynomial wave functions of the linear harmonic oscillator
with the spring constant $\omega^2=8$. This does not mean that from
this information one could immediately deduce the behavior of the
low-lying-state wave functions near $x=0$. The reason is that the
propagation of the errors during the analytic continuation of wave
functions is not under our control.

One of the serendipitous merits of interaction (\ref{pot}) is that
we may easily write down the {\em general} solution of
Eq.~(\ref{SEst}) in the complex vicinity of the origin,
 \be
 \psi_{(c_+,c_-)}(x) \sim c_+\,e^{+g/(2x^2)
 + \ldots} +c_-\,e^{-g/(2x^2) +
 \ldots}\,.
 \label{whichwatch}
 \ee
This formula enables us to see that at $|x| \ll 1$, the two
individual components of these solutions behave differently in the
different small$-|x|$ complex Stokes sectors. The component
$\exp({+g/(2x^2)})$ will be dominant in the right (i.e., ``first")
and left (i.e., ``minus first") rectangular wedges, where $|\arg x|
< \pi/4$ and $|\arg(- x)| < \pi/4$, respectively, and {\em vice
versa}. Although the latter observation may be read just as an $x
\to 1/x$ image of the analogous statements valid for the more common
large$-|x|$ Stokes sectors, the key novelty is that in the present
context the distance of {\em any} acceptable, {\em fixed} complex
integration curve ${\cal C}$ from the point $x=0$ is strictly
greater than zero. The limiting transition to the origin would be a
purely mathematical exercise, therefore.

This being said, it is still interesting to add a few formal
comments on the consequences of the strongly singular character of
our present potential function $V(x)$ at the unphysical point $x=0$.
Firstly, any left plus right branches of a given low-lying wave
function $\psi_n[x(s)]$ may be visualized as matching at $s=0$
(where $x=x(0)=-{\rm i}\varepsilon$, with $\varepsilon>0$ not
necessarily optimal or even large). Secondly, due to the analyticity
of these two functions (which will coincide at the physical
bound-state energy) we may locally deform our straight complex line
$x(s)$ and move the matching point upwards, closer to the origin.
Keeping this deformation of the integration curve left-right
symmetric, we may most simply get arbitrarily close to the origin
along the negative imaginary axis. This means that we shall stay
within the lower, ``zeroth" Stokes' wedge where $|\arg (-{\rm i} x)|
< \pi/4$ so that it will be the minus-sign exponential component
$\exp({-g/(2x^2)})$ which will be growing and dominant in
Eq.~(\ref{whichwatch}).

Clearly, our ${\cal PT}-$symmetric scenario is different from the
current half-line constructions in which one strictly requires that
$c_+=0$ {\em and} in which one approaches the origin from the left
or right, i.e., within the minus first or first small$-x$ Stokes'
sectors. The analytic continuation of these solutions into the
zeroth sector would lead to their unbounded growth near the origin.
Conversely, one might impose the ``anomalous" half-line-like
boundary condition $c_-=0$ in the origin and expect that the
corresponding  new solutions will exhibit the exponential increase
in the ``usual" minus first and first small$-x$ Stokes' sectors.
Such a generalized form of analogy of our present $\gamma=4$ model
with its exactly solvable $\gamma=0$ predecessor (involving, in the
language of energies, the quick decrease of the low lying spectrum
with the growth of the coupling $g$ and/or parameter $R$) would
become only slightly more complicated for the larger exponents
$\gamma$.

\subsection*{Acknowledgement}

Discussions of the subject with Roberto Tateo and with several other
colleagues are gratefully appreciated.

\end{document}